\documentclass[10pt,english]{llncs}
\usepackage[T1]{fontenc}
\usepackage[latin9]{inputenc}
\usepackage{babel}
\usepackage{amsmath}
\usepackage{amssymb}
\usepackage{graphicx}
\usepackage{subfig}

\usepackage{url}
\usepackage{listings}
\usepackage{color}
\usepackage[usenames]{xcolor}
\newcommand*{\xml}[1]{\texttt{<#1>}}

\makeatletter

\begin{document}

\pagestyle{headings}  
\addtocmark{Hamiltonian Mechanics} 

\title{Stealthy Deception Attacks \\ Against SCADA Systems}
\titlerunning{Stealthy Deception Attacks Against SCADA Systems}

\author{Amit Kleinmann\inst{1} \qquad Ori Amichay\inst{1} \qquad {\rm Avishai Wool\inst{1}} \\
   {\rm David Tenenbaum\inst{2}} \qquad {\rm Ofer Bar\inst{2}} \qquad {\rm Leonid Lev\inst{2}}}

\institute{
Cryptography and Network Security Lab, School of Electrical Engineering\\
Tel-Aviv University, Ramat Aviv 6997801, Israel\\
\email{a.b.kleinmann@gmail.com, oriamich@gmail.com, yash@acm.org} \and
Israel Electric Corporation, 1 Netiv Ha'Or Haifa 3100001, Israel\\
\email{\{ david.tenenbaum | br.ofer | leonid.lev \}@iec.co.il}
}

\maketitle
\begin{abstract}
SCADA protocols for Industrial Control Systems (ICS) are vulnerable to network attacks such as session hijacking. Hence, research focuses on network anomaly detection based on meta--data (message sizes, timing, command sequence), or on the state values of the physical process.
In this work we present a class of semantic network-based attacks against SCADA systems that are undetectable by the above mentioned anomaly detection. After hijacking the communication channels between the Human Machine Interface (HMI) and Programmable Logic Controllers (PLCs), our attacks cause the HMI to present a fake view of the industrial process, deceiving the human operator into taking manual actions. Our most advanced attack also manipulates the messages generated by the operator's actions, reversing their semantic meaning while causing the HMI to present a view that is consistent with the attempted human actions. The attacks are totaly stealthy because the message sizes and timing, the command sequences, and the data values of the ICS's state all remain legitimate.

We implemented and tested several attack scenarios in the test lab of our local electric company, against a real HMI and real PLCs, separated by a commercial-grade firewall. We developed a real-time security assessment tool, that can simultaneously manipulate the communication to multiple PLCs and cause the HMI to display a coherent system--wide fake view. Our tool is configured with message-manipulating rules written in an ICS Attack Markup Language (IAML) we designed, which may be of independent interest. Our semantic attacks all successfully fooled the operator and brought the system to states of blackout and possible equipment damage.

\end{abstract}
\textbf{Keywords: }SCADA, Stealthy Deception Attacks, IDS, NIDS, ICS

\section{\label{sec:Introduction}Introduction}
\subsection{Industrial Control Systems (ICS)}
Industrial Control Systems (ICS) are used for monitoring and controlling numerous industrial systems and processes. In particular, ICS are used in critical infrastructure assets such as chemical plants, electric power generation, transmission and distribution systems, water distribution networks, and waste water treatment facilities. ICS have a strategic significance due to the potentially serious consequences of a fault or malfunction.

ICS typically incorporate sensors and actuators that are controlled by PLCs, and which are themselves managed by the HMI. PLCs are computer-based devices that were originally designed to perform the logic functions executed by electromechanical hardware (relays, switches, mechanical timer, and mechanical counters). An automation system within a campus is usually referred to as a Distributed Control Systems (DCS), while Supervisory Control And Data Acquisition (SCADA) system is a type of ICS that typically comprises of different stations distributed over large geographical areas.

Most SCADA network traffic is generated by automated processes and mainly for data acquisition, in the form of periodic polling of field devices. The control is done by commands that are used to change the operation state of the PLC and/or its controlled equipment, e.g., a circuit switch. Provision against unauthorized actions are system- or process-dependent.

ICS were originally designed for serial communications, and were built on the premise that all the operating entities would be legitimate, properly installed, perform the intended logic and follow the protocol. Thus, many ICS have almost no measures for defending against deliberate attacks. Specifically, ICS network components do not verify the identity and permissions of other components with which they interact (i.e., no authentication and authorization mechanisms); they do not verify message content and legitimacy (i.e., no data integrity checks); and all the data sent over the network is in plaintext (i.e., no encryption to preserve confidentiality). Therefore, ICS networks are vulnerable to cyber attacks, and in particular to session--hijacking attacks.

Anomaly--based intrusion detection approaches are based on ``the belief that an intruder's behavior will be noticeably different from that of a legitimate user'' \cite{mukherjee1994network}. The main types of anomaly detection approaches that are applied to SCADA systems \cite{alcaraz_cazorla_fernandez,atassi,chen} are: Network--aware detection in which the anomaly detection models only consider network and OS-level events; Protocol--aware detection in which modeling the normal traffic relies on deep-packet-inspection and considers the SCADA control protocol's meta-data (message sizes, timing, argument addresses, command sequence); and Process--aware approaches which are based on process invariants, mathematical relationships among physical properties of the process controlled by the PLCs.

\subsection{Contributions}
In this work we present a class of semantic network--based attacks against SCADA systems, that are undetectable by either protocol--aware or process--aware anomaly detection. After hijacking the communication channels between the HMI and PLCs, our attacks manipulate the traffic so as to cause the HMI to present a fake view of the industrial process, thus deceiving the human operator into taking inappropriate and damaging manual actions. Our most advanced attack also manipulates the messages generated by the operator's actions, reversing the semantic meaning of commands (`Close' becomes `Open' and vice-versa) while causing the HMI to present a view that is consistent with the attempted human actions---thus inducing real damage on the cyber-physical system.

Our attacks are totaly stealthy to SCADA-aware anomaly detectors since the message sizes and timing, and also the command sequences (including the command arguments) are all 100\% legitimate in every way. Furthermore, our attacks are undetectable even by process-aware anomaly detection, since the observed data values of the ICS's state are completely legitimate: they appear as natural fault conditions that the SCADA system is designed for, and the human operator is trained to handle. Even the operator's manual command sequences are the expected actions when responding to a natural fault.

We implemented and tested several attack scenarios in the test lab of our local electric company, against a real HMI and real PLCs, separated by a commercial-grade firewall. To do so we developed a real-time security assessment tool, that can simultaneously manipulate the Modbus communication between the HMI and multiple PLCs, and cause the HMI to display a coherent system--wide fake view.

Our tool is configured with message-manipulating rules written in an ICS Attack Markup Language (IAML) we designed. Our tool is near-stateless---it only replaces message contents, without injecting, deleting, extending, or shortening messages. In one scenario we even used the tool in a ``half-duplex'' mode, wherein it only manipulates the HMI-to-PLC queries, while the PLC-to-HMI responses are unaltered. Despite these self--imposed limitations, our multi-stage semantic attacks all successfully fooled the operator and brought the system to states of blackout and possible equipment damage.

The rest of this paper is organized as follows: Section 2 explains some fundamentals and assumptions.  Section 3 describes the security assessment tool. In Sction 4 we present three attacks we tested with their impact on the HMI and the PLCs. Section 5 describes the ICS Attack Markup Language (IAML). Section 6 succinctly reviews related work and in Section 7, we suggest possible countermeasures and conclusions. Examples of IAML can be found in an Appendix.

\section{Preliminaries}
\subsection{Modbus}
Modbus is a de facto standard for ICS. Many Modbus systems implement the communications layer using TCP as described in the Modbus over TCP/IP specification \cite{modbus2006modbus}. The specification defines an embedding of Modbus packets in TCP segments. TCP port 502 is reserved for Modbus communications. The Modbus protocol employs a simple master-slave communication mode. The master device initiates transactions (called queries) and the slaves respond by supplying the requested data to the master or by performing the action requested in the query. Only one device can be designated as the master (usually the HMI) while the remaining devices are slaves (usually PLCs). A slave sends a response message for every query that is addressed to it. A unique transaction ID is created for the request message from the master, which the slave includes in its response.

Each PLC provides an interface based on the Modbus data model. The data model is comprised of ``coil'' (single-bit) and ``register'' (16-bit) tables. Read and write operations associated with these items can access multiple consecutive data items. The Modbus PDU has two fields that refer to the data model: a single-byte {\em Function code} and a variable size {\em Payload} (limited to 252 bytes), which contains parameters that are specific to the function code. A read request payload has two fields, a reference number and bit/word count. The reference number field specifies the starting memory address for the read operation. The bit/word count field specifies the number of memory object units to be read. The payload of the corresponding response has two slightly different fields, byte count and data. The byte count specifies the length of the data in bytes. The data field contains the values of the memory objects that were read. In addition to memory references, the payload of a write message has fields that specify the values that are to be written.

The Modbus protocol does not defend itself in any way against a rogue master that sends commands to slaves. Furthermore, Modbus only relies on TCP sequence numbers to provide session semantics and has no message integrity defences, thus TCP session hijacking \cite{bellovin1989security} is quite straightforward.

Modbus over TCP/IP has long-term session semantics -- the protocol simply involves separate two-message  query-response sequences. However, the Unitronics PLCs we tested only accept a single TCP connection at a time on port 502. Therefore, an attacker attempting to control an already-controlled PLC would need to either hijack the existing TCP connection \cite{bellovin1989security} and inject spoofed packets into the stream, or reset the existing connection and create a new connection. PLCs that allow multiple concurrent connections on port 502 are susceptible to much simpler attacks.

\subsection{Electrical Distribution}
\begin{figure*}[t]
\centering
               \includegraphics[
                     width= 0.97
                     \textwidth, 
                      trim = -10 50 0 75
               ] {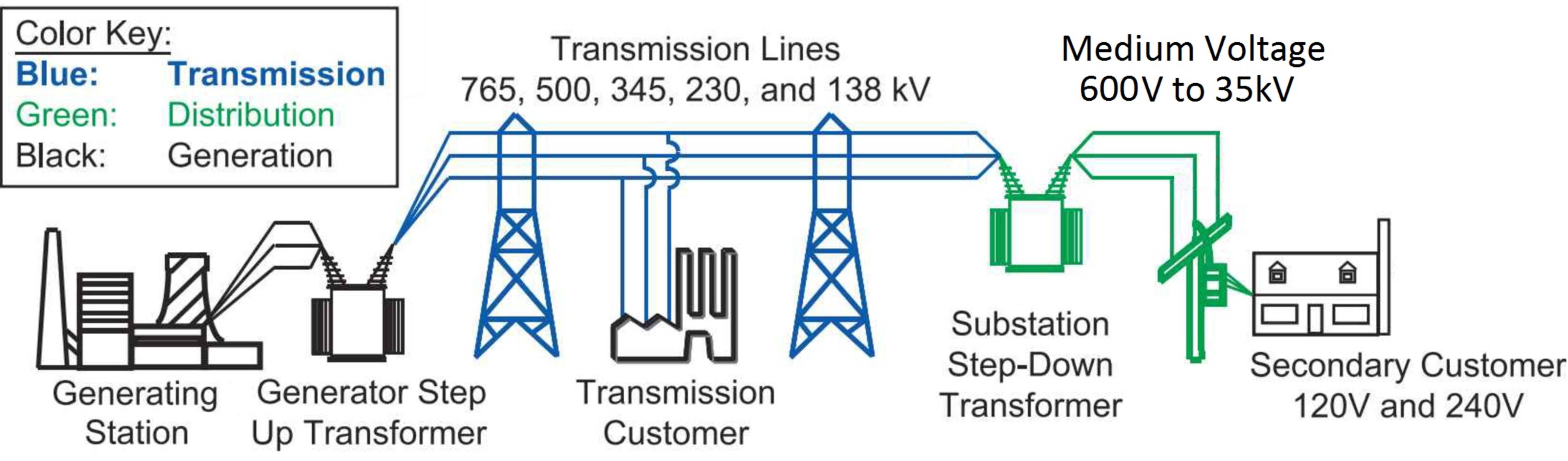}
  \caption{ Basic Structure of the Electric System (following \cite{OutageTaskForce})}
  \label{fig:ElectricityNetwork}
\end{figure*}

An electricity supply chain is usually divided into three subsystems: generation, transmission, and distribution, as depicted in Figure  \ref{fig:ElectricityNetwork}. Electricity is transported along high voltage transmission lines (the transmission network) over long distances, from generation sites to major distribution points. The transmission lines are connected to distribution substations. At a distribution substation, a substation transformer takes the incoming transmission-level voltage (138 to 765 kV) and steps it down to several distribution primary circuits (``medium-voltage'' circuits, 600V to 35kV), which fan out from the substation. Close to each end user, a distribution transformer takes the medium-voltage and steps it further down to a low-voltage secondary circuit (commonly 120/240V).
In this paper we focus on the distribution subsystem, between the substation and distribution transformers---which is precisely the subsystem impacted during the Ukranian cyber-attacks \cite{lac-sans-2016,liang20162015}.

For improved reliability, distribution circuits are often provided with ``tie switches'' to other circuits which are normally open (i.e., disconnected). If a fault occurs on one of the circuits, the tie switches can be closed (connected) to let electricity flow into the faulted circuit, and to allow some portion of the service to be restored. The tie switches can be operated either manually, or automatically from the SCADA system interface. These switches, also called switchgears, may be simple open-air isolator switches or may be gas-insulated.

\subsection{Adversary Model}
Our underlying threat model is loosely based on the Dolev-Yao threat model \cite{Dolev:1981:SPK:891726}: The adversary may overhear and intercept all traffic regardless of its source and destination. More precisely, we assume the adversary has a Man-In-The-Middle (MITM) position between the HMI and {\it all} the PLCs. The adversary can inject, delete, and delay arbitrary packets with any source and destination addresses on the communication channels it controls. Consequently, the adversary can also replay previously overheard messages, or maniplulate messages in transit. In particular the adversary can take over the HMI and issue control messages. The objective of the adversary is to manipulate the SCADA network to achieve an impact on the physical world.

We further assume that the adversary has in-depth knowledge of the architecture of the SCADA network and the various PLCs as well as sufficient knowledge of the physical process and the means to manipulate it via the SCADA system. Thus the adversary has the ability to fabricate messages that would result in real-world physical damage.

In our experiments we implemented a somewhat weaker type of attacker: our attack system has network access between the HMI and PLCs, and implements simultaneous MITM attacks against multiple HMI-PLC communication channels. However, Our attack tool is near-stateless and does not track or modify the TCP sequence numbers. Hence our attacks do not inject fabricated messages or drop legitimate ones: our attack tool only modifies the contents of pre-existing messages. Despite this self imposed restriction, our attacks are all successful, and undetectable by suggested anomaly--detection systems.

\begin{figure*}[t]
\centering
               \includegraphics[
                     width= 0.97
                     \textwidth,
                      trim = 0 20 0 30
               ] {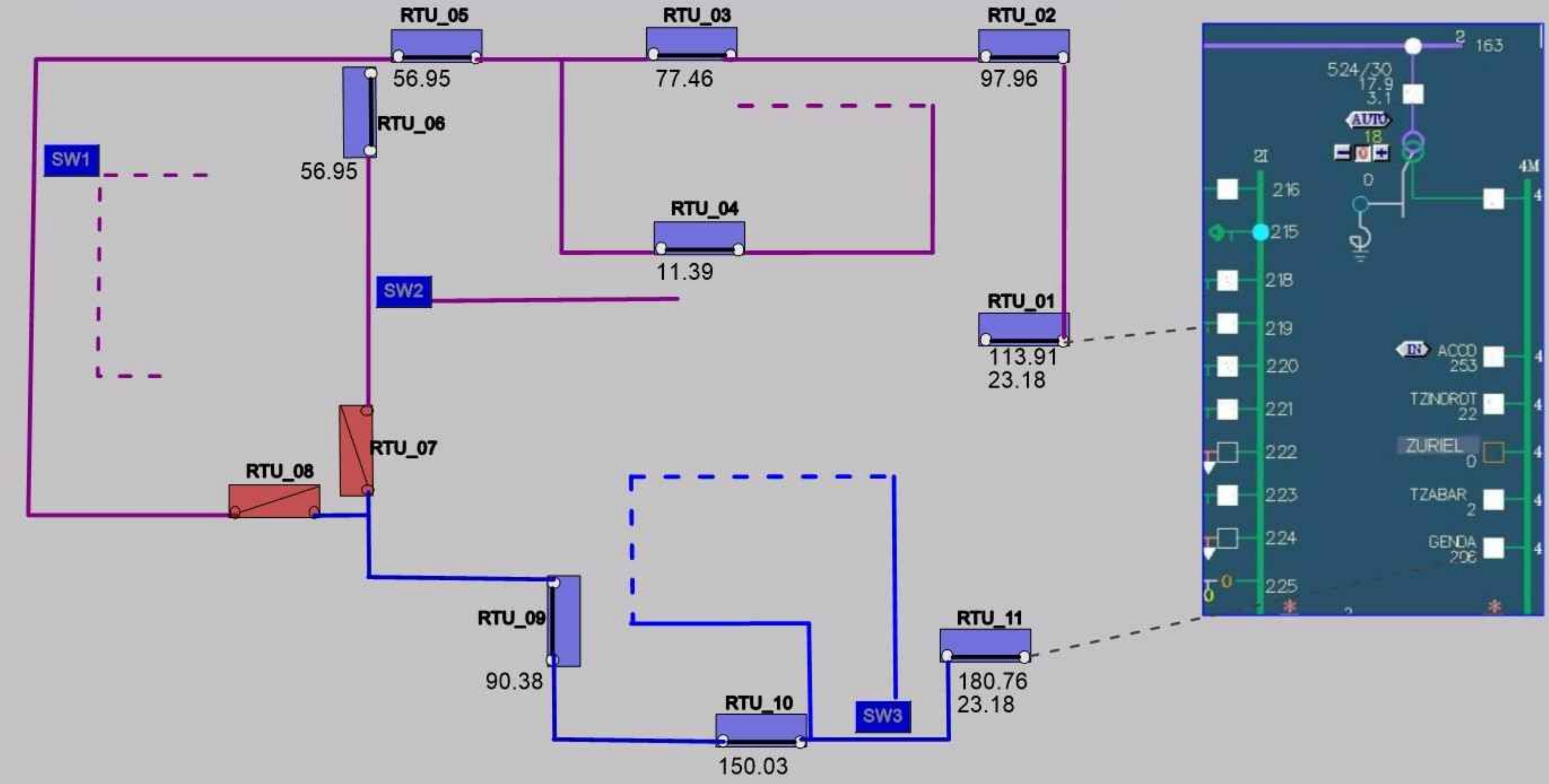}
  \caption{A screenshot of the HMI panel. We see the substation is on the right, and 2 radial distribution lines: the top line (purple, via RTU 01--06) and the bottom line (blue, via RTU 11,10, 09). RTU 07 and 08 control the switchgears that tie the two lines: both are shown as open (disconnected). The numbers below each RTU show the current (113.91A and 180.76A at the line heads, dropping to 11.39A at RTU 04). At the two line heads (RTU 01, 11) the HMI also shows the voltage (23.18KV) displayed.}
  \label{fig:sim-panel}
\end{figure*}

\begin{figure}[t]
\centering
  \includegraphics[ 
  width= 0.6
   \textwidth, 
   trim = 30 100 0 50
   ]{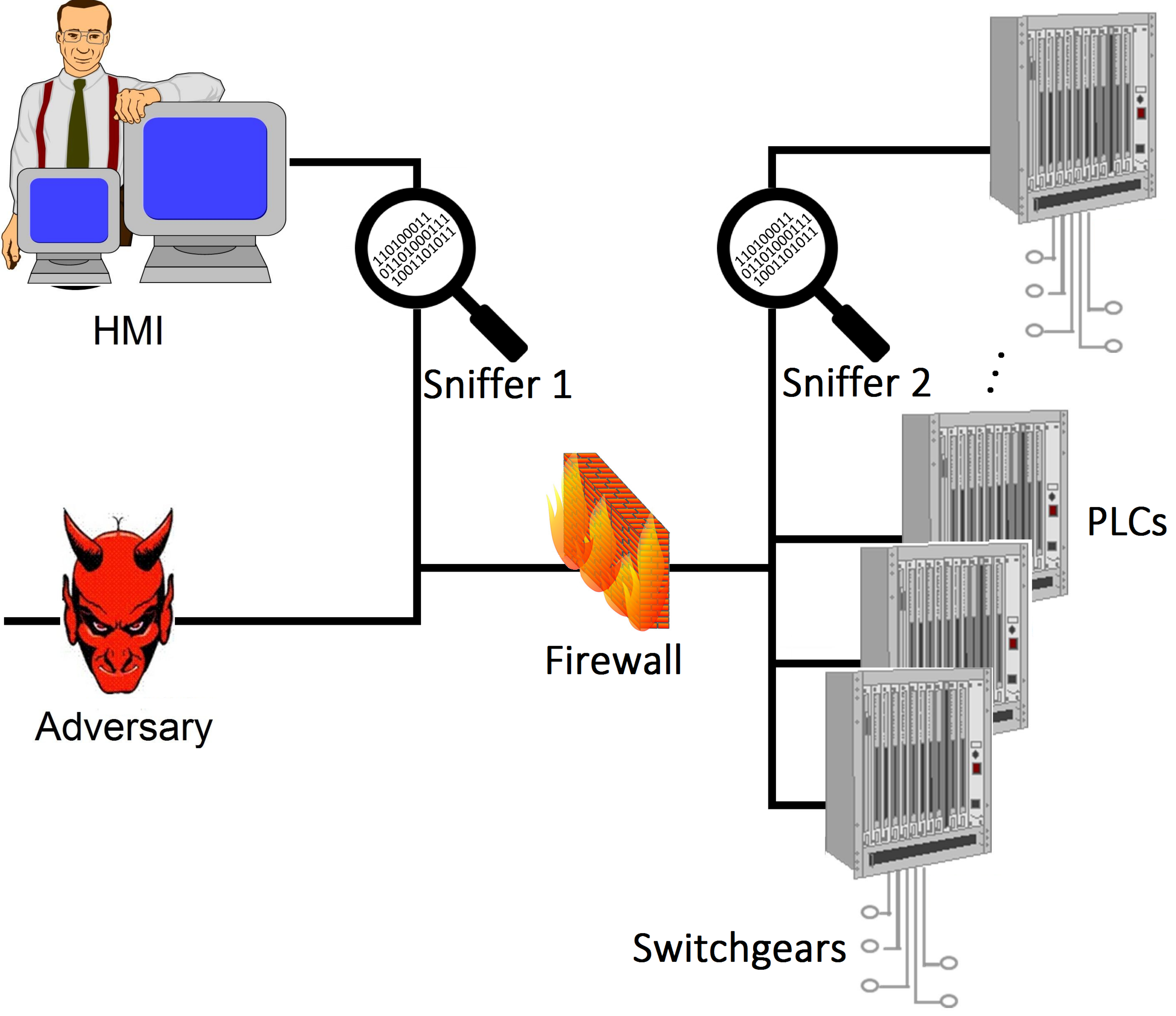}
   \hfill
\caption{Network diagram of the test lab}
    \label{fig:attackerAccess}
\end{figure}

\subsection{The Test Lab}
The electric company test lab consists of 11 Unitronics V130 PLCs, controlled by a Cimplicity version 9.5 HMI running on a Microsoft Windows 2012 server. The HMI and the PLCs are connected to separate VLANs and separated by Check Point 4000 appliance R77.30 firewall.

The test lab emulates a substation with 2 radial distribution lines, that are interconnected by tie switches. Moreover, along each line there are several additional switchgears, see Figure \ref{fig:sim-panel}. In total there are 11 PLCs, each controlling a switchgear: each PLC reports back to the HMI the voltage and current flowing through it, and the switchgear state (open/closed), and accepts commands to open or close the switchgear.

The distribution system controlled by the PLCs is simulated by a separate PLC (the S-PLC) - which the switchgear PLCs interface with. Whenever a switch attempts to read a sensor value (e.g., current), the S-PLC provides the required value. The sensor values reported by the S-PLC are based on measurements taken at real switchgears that are deployed at a certain radial circuit of 
our local electric company's network.

The switchgears that are controlled by RTU\_07 and RTU\_08 in Figure \ref{fig:sim-panel} are the tie switches and are initially {\it disconnected}. The initial state of all the other switchgears is {\it connected}.

The HMI runs two separate polling threads that monitor the eleven PLCs using the Modbus protocol. The polling threads repeatedly send the same three read-requests to get register values from each one of the PLCs. The PLCs all run the same control logic and expose the same Modbus memory layout: in particular the current value is stored in register \#130 and the voltage value is stored in register \#131.

In this environment, and indeed in the electric company's real HMI, switchgears are only operated manually from the HMI: If the operator observes a fault, such as current and voltage dropping to zero, she can open or close the switchgears by clicking on the HMI screen. Such operator actions are realized by corresponding Modbus `write' packets that are sent from the HMI to the proper PLC. Note that in the test lab, the S-PLC reacts to such write events by updating all the subsequent current and voltage values that will be reported to all the relevant PLCs to be consistent with the system state following the operator action.

\section{The Attack Tool}
\subsection{Gaining Network Access}
There are many ways for an attacker with network access to place itself in a MITM position. In our attack tool we chose to implement a well known ARP poisoning attack (cf. \cite{abad2007analysis}). Using ARP poisoning obviously makes the attack tool detectable to standard low level Network Intrusion Detection Systems (NIDS). However, our focus is on semantic SCADA-aware anomaly detection, so we assume the attacker is able to bypass the NIDS somehow. In our experiments we realized that the HMI occasionally receives ARP messages, triggered by other network connections, causing its ARP table to recover. Hence the attack tool has to keep sending the spoofed ARP messages (in our case, once every 2 seconds was sufficient).

In order to record the attacks' progress, we placed 2 traffic sniffers, one on each of the VLANs (Sniffer 1 and Sniffer 2 in Figure \ref{fig:attackerAccess}).

\begin{figure}[t]
\centering
  \includegraphics[ 
   width= 0.7
   \textwidth, 
   trim = -75 100 0 100
   ]{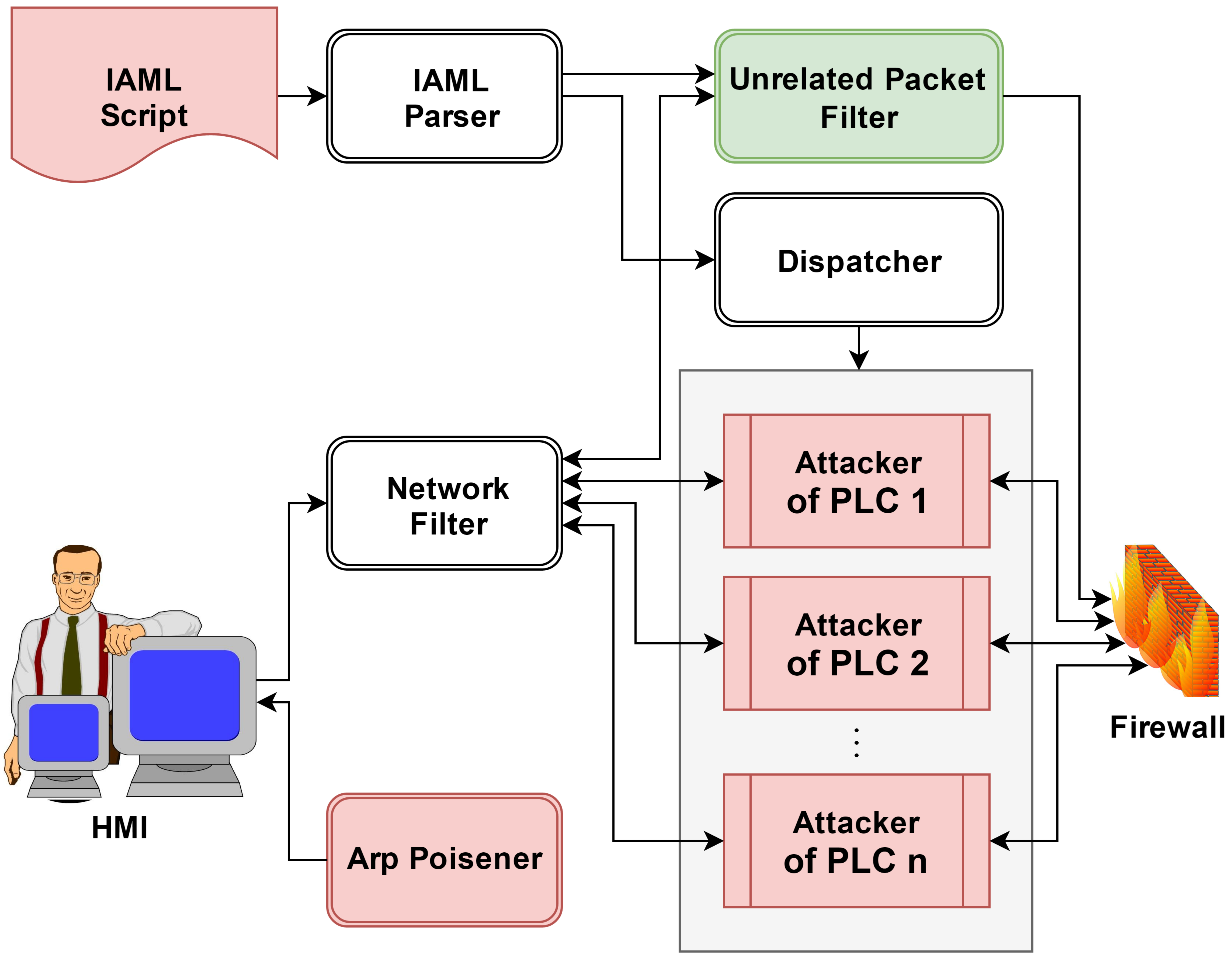}
   \hfill
\caption{The architecture of the attack tool}
    \label{fig:attackerTool}
\end{figure}

The architecture of the attack tool is depicted in Figure \ref{fig:attackerTool}. It comprises of: an {\it Arp poisener} that crafts and sends the spoofed-ARP messages to the HMI, an {\it Unrelated Packet Filter (UPF)} that just forwards these packets (untouched) to their original destination, a {\it Dispatcher} that creates a {\it PLC attacker} (composed of 2 threads, sharing state, per each direction of the traffic) per each PLC that needs to be attacked. An {\it IAML parser} that accepts an IAML script, parses it and transfers the needed parameters to the {\it Dispatcher} and the {\it UPF}, and a {\it Network filter} that filters proper packets from the HMI VLAN and sends them to the appropriate {\it PLC attacker}. The Network filter is based on the Pcap.Net .NET wrapper for WinPcap.

\subsection{Near-Stateless Manipulation of Modbus}
Modifying the message length in the Modbus stream is a relatively ``noisy'' attack action. SCADA-aware anomaly detection that has even minimal Modbus understanding can flag messages with unusual lengths. Further, injecting messages into a hijacked connection is also detectable by either a network--aware approach (if the message timing is unusual) or by a protocol--aware approach (if the function code or arguments are unusual). Therefore, to demonstrate the power of our attacks we elected to make our attack tool stateless at the transport layer: it does not track or modify the TCP sequence numbers at all. Note that this introduces a significant limitation on the attacker: e.g., it precludes replacing a Modbus ``read'' query by a ``write'' query (which is the usual way to implement commands)---simply because a ``write'' message is longer, due to the additional written-values payload.

However our tool is not totally stateless: As we shell see, we wish to modify the values reported to the HMI in PLC responses of selected messages.  Importantly, the Modbus response messages do not carry the read register addresses; they carry only the read data values, while the register addresses are present in the HMI's query message. Thus when the attack tool matches a query message of interest, it places that query's Modbus Transaction ID (TID) into a state variable; when the corresponding response message, with the same TID, is matched on the same connection, the attack tool modifies its content.

\begin{figure*}[t]
 \centering
\subfloat[Effects of the attack as measured at the HMI
             \label{figure:attack0-hmi}
            ]{
               \includegraphics[
                     width= 0.96
                     \textwidth,
                        trim = 10 5 10 30
               ] {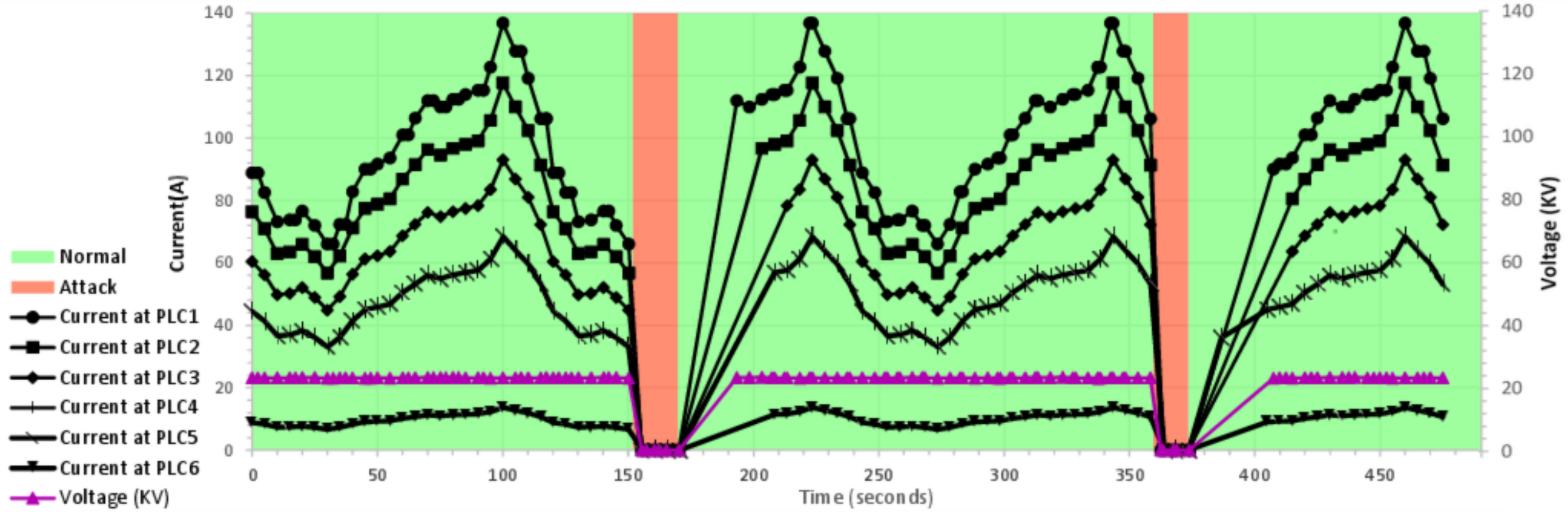}
             } \\[-2ex]
\subfloat[Actual values during the attack as reported by the PLCs
             \label{figure:attack0-plc}
            ]{
               \includegraphics[
                     width= 0.96
                     \textwidth,
                     trim = 10 5 10 -30
               ] {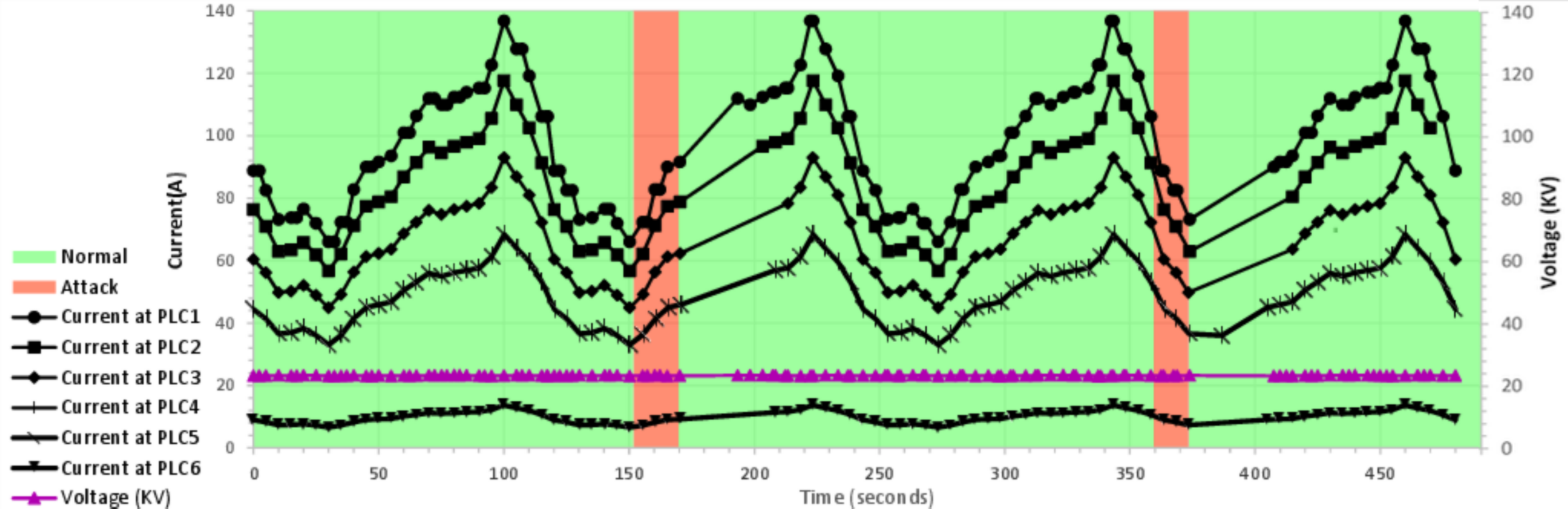}
             }
  \caption{The {\em zero values deception attack} against an ICS of an electric distribution system. The attack sends zero values for the current and voltage reported simultaneously by RTU\_01 - RTU\_06 (denoted here as PLC1--PLC6 respectively), causing the HMI to present a view consistent with a fault on the top line.}
  \label{fig:attack2}
\end{figure*}

\section{Semantic Attacks against Electric Distribution SCADA} 
\subsection{Zero Values Deception Attack} \label{ZeroValuesAttack}
Our first deception scenario is illustrated in  Figure \ref{fig:attack2}. The attacker invokes the attack twice (first for 18 seconds between 151.8 - 169.8 sec, and second for 14 seconds, between 359.53 - 373.53 sec). During each one of these attacks the attacker simultaneously changes the values of the registers (\#130 and \#131) that hold the current and the voltage reported by six PLCs to be zeros. The victim PLCs are those controlling the top line (recall Figure \ref{fig:sim-panel}): RTU 01--06. The bottom graph of Figure \ref{fig:attack2} shows the true current and voltage values as recorded by Sniffer 2 on the PLC VLAN. The top graph shows the values of the same registers as recorded by Sniffer 1 and observed at the HMI.

This attack causes the HMI to display a view in which RTU\_01 - RTU\_06 all show zero current - as in Figure \ref{fig:sim-panelDisconnect}. This view is consistent with a natural fault on the top line -- and causes the operator to implement unneeded remediation steps, that are expensive and possibly harmful. Note that the attack is super-stealthy: SCADA-aware anomaly detection is blind to such an attack, since the attack mimics a natural fault, which is a planned-for scenario and does not, in itself, signify an attack.

\subsection{A Multi-stage Attack}
\begin{figure*}[t]
\centering
               \includegraphics[
                     width= 0.97
                     \textwidth,
                      trim = 0 20 0 70
               ] {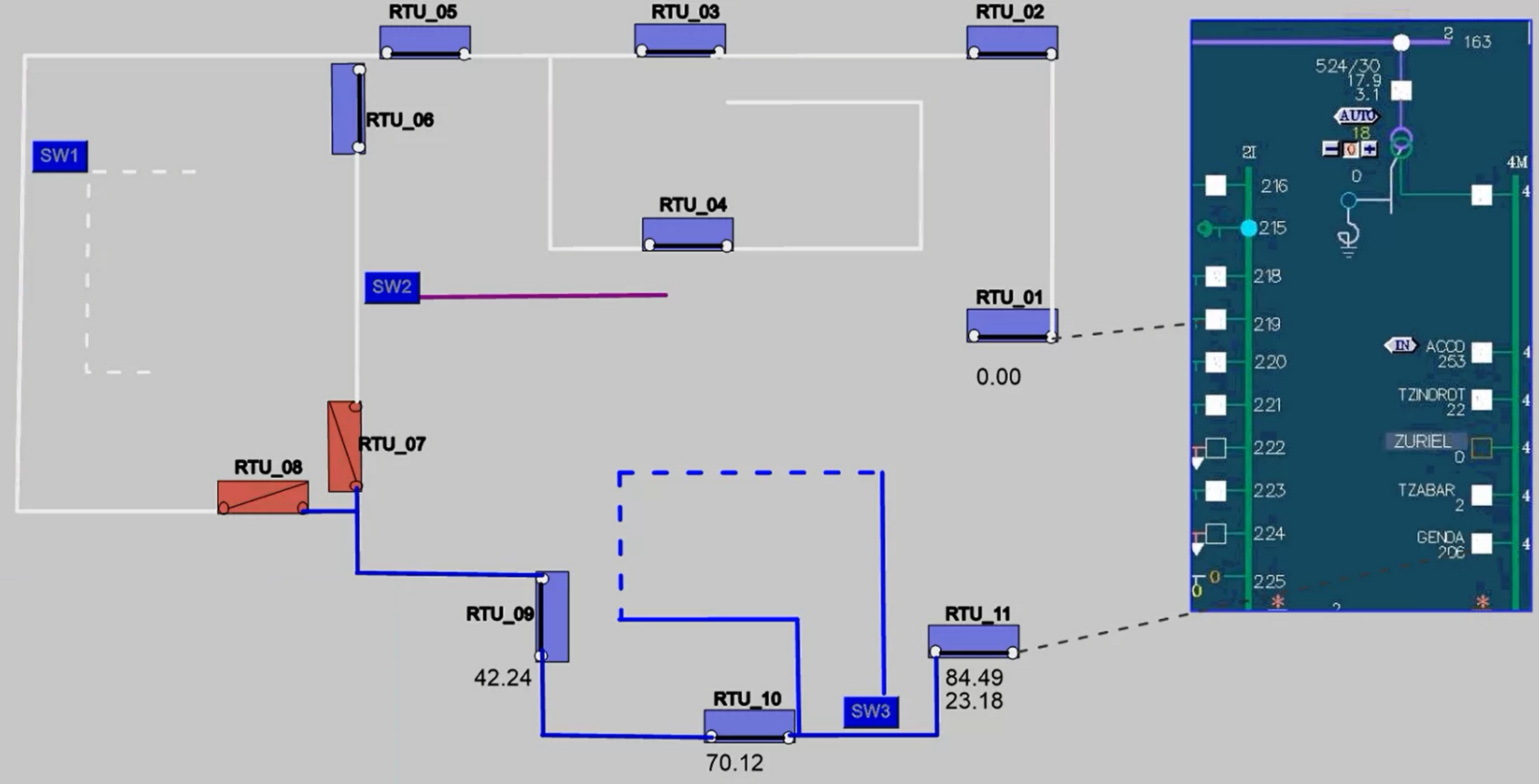}
  \caption{The HMI panel showing a disconnection in the top radial line. Note the voltage of 0.0KV at RTU\_01, the missing current value at RTUs 02--06, and the white color of the line. In such a scenario the operator would typically open the switchgear at RTU 01 and close the ties at RTU 07, 08 to  attempt to supply the top line from the bottom line.}
  \label{fig:sim-panelDisconnect}
\end{figure*}

\begin{figure*}[t]
\centering
\subfloat[Effects of the attack as measured at the HMI
             \label{figure:attack2-hmi}
            ]{
               \includegraphics[
                     width= 0.96
                     \textwidth,
                      trim = 10 5 10 30
               ] {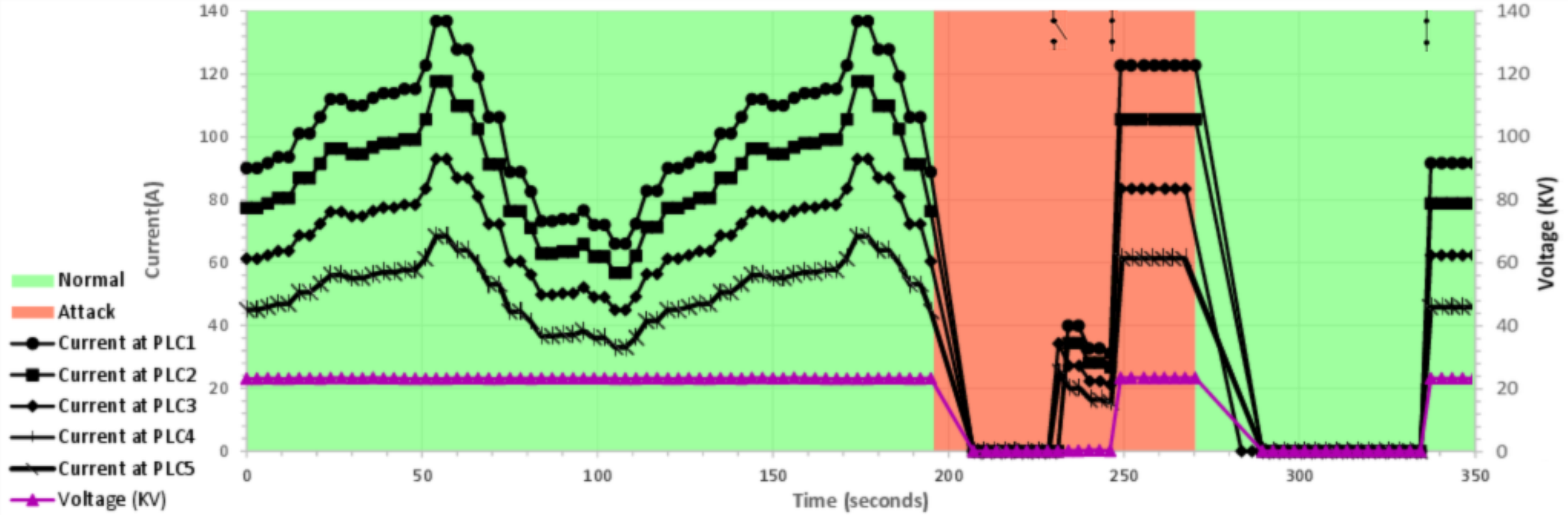}
             } \\[-2ex]
\subfloat[Actual values during the attack as reported by the PLCs
             \label{figure:attack2-plc}
            ]{
               \includegraphics[
                     width= 0.96
                     \textwidth,
                     trim = 10 5 10 -20
               ] {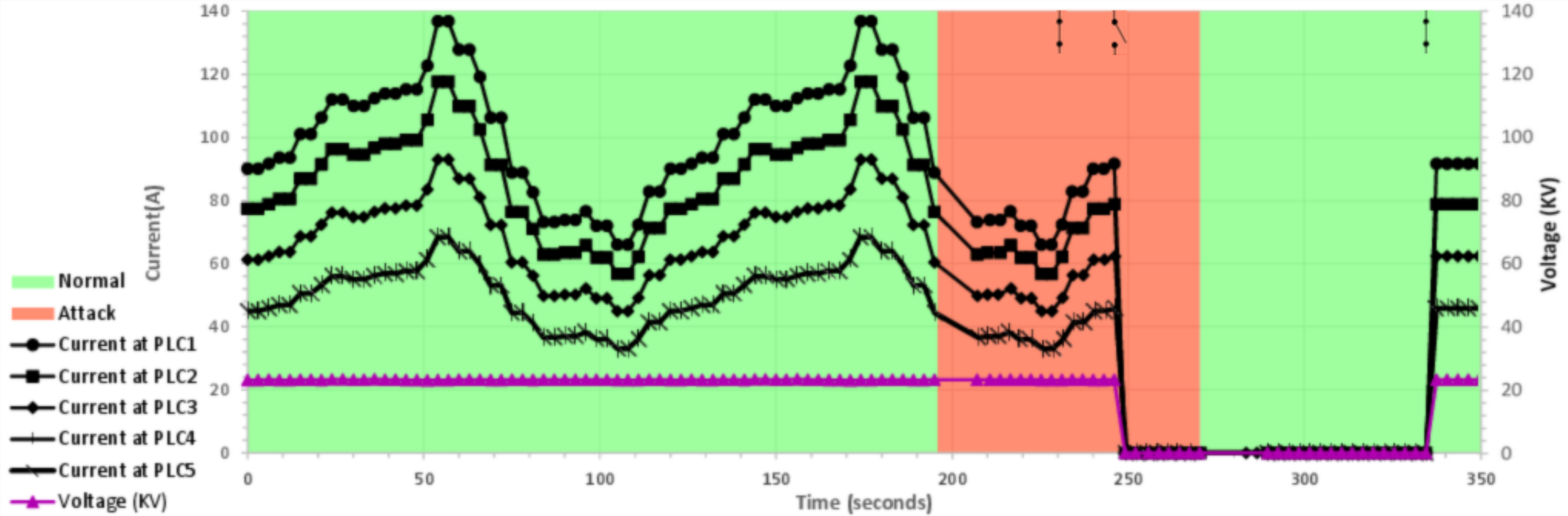}
             }
  \caption{The multi-stage attack -- the attack starts at time 195.8 with a zero-values attack. The second stage executing the ``opposite operation'' attack starts at 231.3 and ends at 270.4, with operator commands at times 231.3 and 246.6. The icons at the top edge of the graphs indicate the {\it Open} and {\it Close} commands to the switchgears.}
  \label{fig:attack3}
\end{figure*}

Our main attack is more elaborate, and aims to interfere with the operator's reaction to a fake fault. In this attack scenario we implemented a stealthy multi-stage deception attack, see Figure \ref{fig:attack3}. The attack has two main stages: the first stage is the Zero values deception attack described in the previous section. Looking at the HMI pannel (depicted at Figure \ref{fig:sim-panelDisconnect}) the operator is deceived to believe that the top radial circuit is disconnected. This motivates the operator to start disconnecting and reconnecting switchgears according to the operating procedures. This is where the second stage of the attack comes into action. In this stage, whenever the operator issues a switchgear open/close command, the attack tool replaces it with the {\it opposite command}.

The bottom graph of Figure \ref{fig:attack3} shows the actual values (of current and voltage) recorded at the different PLCs, and the top graph shows the manipulated values observed at the HMI. The two graphs also indicate the open/close actions by small icons: the top graph shows the intended operator actions at the HMI ({\it Open} at time 231.3 and {\it Close} at time 246.6), and the bottom graph shows the icons for the reversed action received and executed by the PLCs. Note that the attack also fakes current and voltage values that are consistent with the intended operator actions: after the {\it Open} command the attack starts returning current values computed as ``nominal $ - $ actual'', where ``nominal'' is a fixed per PLC constant; this shows the operator that the trouble shooting has some effect. Then after the {\it Close} command (at time 246.6) the attack tool reports ``nominal'' current values on all the PLCs---while in fact the circuit is disconnected and customers are experiencing a blackout. Again, note how stealthy the attack is: all Modbus messages are on schedule, using normal functions and arguments, with designed-for, semantically reasonable, data values.

\subsection{A Half-duplex Attack}
Recall that the zero-values attack of section \ref{ZeroValuesAttack} required the attacker to modify values in responses from the PLC to the HMI. In this section we describe an attack that is equivalent to the zero-values attack, except it requires a simpler setup by the attacker: Here the attacker acts as a MITM only on the traffic that flows from the HMI to all the PLCs while the PLC-to-HMI responses are unaltered. Hence the attacker is unable to directly set data values to zero in PLC responses. To achieve an attack in this scenario, query messages (from the HMI to the PLC) reading the current and voltage values are modified to access registers whose addresses are shifted by 1. As a result, during the attack, the shifted register-values sent by the PLCs are interpreted at the HMI as the values of the corresponding preceding registers---the HMI interprets the voltage value as the current, and the content of register \#132 (which is 0) is interpreted as voltage-value, again fooling the operator into deducing that a fault occurred. We omit the graphs.

Note that this attack can be detected by a protocol-aware anomaly detection \cite{Goldenberg201363,Kleinmann:2016:ACS:2994487.2994490} as long as the detector is located at the network segment where the PLCs reside, since the attacker modifies the accessed register addresses, which creates unknown symbols in the GW model \cite{Goldenberg201363}.

\section{The ICS Attack Markup Language} \label{IAML}
We defined a new formalism for specifying a concrete execution of an attack, an ICS Attack Markup Language which we called an IAML. IAML enables planning and implementation of a multi-stage, multi-PLC, simultaneous attacks on an ICS without any a priori programming knowledge. It uses a modular approach, where a module represents an attack that changes a certain type of packet under certain conditions. Modules can be linked together to create multi-stage attacks. The relationships among modules are specified through the conditions and changes of local and global stages. IAML is accompanied by a library of predicates, which function as a vocabulary to describe the properties of attack modules and stages.

\subsection{Syntax and Elements}
In order to configure the attack tool in a data-driven manner, we designed the ICS Attack Markup Language (IAML). IAML is written in XML -- Listing \ref{lst:IamlAttackMultistage1} in the appendix shows the IAML syntax with the basic elements. The root \xml{IAML} tag has one sub-element: \xml{Change}.

A \xml{Change} tag is the basic unit for specifying changes to a SCADA packet. The \xml{Change} tag accepts one attribute: {\it PacketToChange} (``REQUEST'' if the packet to be changed is a request packet, ``RESPONSE'' in case a response packet needs to be changed). Note that a \xml{Change} of the ``RESPONSE'' may actually match the query, and later trigger a modification of the corresponding response. The \xml{Change} tag has two sub-elements: \xml{Query} and \xml{NewValues}.

The \xml{Query} element supplies the match criteria identifying the packet to manipulate. \xml{Query} has one sub-element \xml{QueryEntry}. \xml{QueryEntry} accepts a pair of attributes: {\it Key} and {\it Value}. One QueryEntry is mandatory: its {\it Key} is ``TYPE'' and its {\it Value} is either ``REQUEST'' or ``RESPONSE'' (it can be set to ``REQUEST'' even when the {\it PacketToChange} is of ``RESPONSE'' type, in case the match criteria is given using the request arguments).

Other optional \xml{QueryEntry}s are the following: \newline
``PLC\_IP'', ``GLOBAL\_STAGE'', ``LOCAL\_STAGE'', ``FUNCTION'', \newline
``WORD\_COUNT'', and ``ADDRESS''. \newline
 The values corresponding to these keys are:
 \begin{itemize}
 \setlength\itemsep{0.15 em}
\item ``PLC\_IP'' - is the IP address of a PLC.
\item ``GLOBAL\_STAGE'' - an integer representing the current global stage of the attack. This is needed, for instance, to synchronize the move from the {\it zero-values deception} stage to the {\it opposite command} stage of the exemplified multi-stage attack.
\item ``LOCAL\_STAGE'' - an integer denoting the current PLC-local stage of the attack. This is needed for finer manipulation on each PLC separately.
\item ``FUNCTION'' - the original function code of the SCADA packet to be matched.
\item  ``WORD\_COUNT'' - the word count in the SCADA packet, e.g., in the case of a Modbus read request - the number of words to read.
\item ``ADDRESS'' - the register number to be matched. Note that the Modbus request may read a range of several registers, and the ``ADDRESS'' may refer to a register inside the range, e.g., if the packet specifies a read of 4 words starting from address \#129, then address \#130 is within this range.
\end{itemize}

The \xml{NewValues} element specifies the change to be made to the SCADA packet. \xml{NewValues} has one sub-element \xml{NewValueEntry}, that accepts a pair of attributes: {\it Key} and {\it Value}. {\it Key} may be equal to: ``GLOBAL\_STAGE'', ``LOCAL\_STAGE'', ``FUNCTION'', ``STARTING\_ADDRESS'', or ``DATA''.
Some of these keys are already described above. The others are:
\begin{itemize}
\item ``STARTING\_ADDRESS'' - the address explicitly specified in the SCADA packet.
\item ``DATA'' - the value to be placed inside the SCADA packet.
\end{itemize}

IAML allows setting the new value as a simple mathematical expression, which is allowed to refer to constants and to the actually-read values of that register:
These values will be the new values that would be written in the modified SCADA packet. If multiple values are specified they should be separated by commas.
The original value is denoted by 'X'. Hence, the new value can be an arithmetic expression, based on 'X', such as: 'X+5'.
The language allows for multiple new-values (or arithmetic expressions) each replacing a corresponding original value and separated by a comma.

\section{Related Work}
\subsection{Attacks against ICS}
Digital attacks that cause physical destruction of equipment do occur \cite{gorman2016}. Most recently, cyber-attacks on SCADA systems controlling electrical distribution have caused wide-spread blackouts in Ukraine \cite{lac-sans-2016,liang20162015}. Perhaps most well known is the attack on an Iranian nuclear facility in 2010 (Stuxnet) to sabotage centrifuges at a uranium enrichment plant. The Stuxnet malware \cite{falliere2011w32,langner2011stuxnet} worked by changing centrifuge operating parameters in a pattern that damaged the equipment -- while sending normal status messages to the HMI. In 2014, the German Federal Office for Information Security announced a cyber attack at an unnamed German steel mill, where hackers manipulated and disrupted control systems to such a degree that a blast furnace could not be properly shut down, resulting in ``massive''-though unspecified-damage \cite{de2014lage}.

Byres et al. \cite{byres2004use} describe the application of the attack tree methodology to SCADA communication systems based on the common Modbus protocol stack. The authors identify eleven possible attacker goals with their respective technical difficulty, possible severity of impact and likelihood of detection. In particular they noted that an attacker can perform a Man-In-The-Middle (MITM) attack between a PLC and HMI and ``feed'' the HMI with misleading data, allegedly coming from the PLC -- which is what we implemented.

In 2009 Fovino et al. \cite{NaiFovino2009139} showed that malware was able to disrupt or even seize control of vital sensors and actuators. Semantic attack scenarios on ICSs are described by \cite{fovino2010modbus,RobertT} for a system with a pipe in which high pressure steam or fluid flows. The pressure is regulated by two valves. An attacker capable of sending packets to the PLCs can force one valve to complete closure, and force the other to open. Each of these ICS commands is perfectly legal when considered individually, however when sent in an abnormal order they can cause a `water hammer' and bring the system to a critical state. Another example \cite{RobertT} shows an attack scenario where a system-wide water hammer effect is caused. A fluid in motion is forced to stop or change direction suddenly, resulting in pressure surge or wave propagation in the pipe. The water hammer is caused simply by opening or closing major control valves too rapidly. This can result in a large number of simultaneous main breaks.

\subsection{Anomaly Detection in ICS}
Surveys of techniques related to learning and detection of anomalies in critical control systems can be found in \cite{alcaraz_cazorla_fernandez,atassi,chen}.
While most of the current commercial network intrusion detection systems (NIDS) are signature-based, i.e., they recognize an attack when it matches a previously defined signature, anomaly-based NIDS
``are based on the belief that an intruder's behavior will be noticeably different from that of a legitimate user'' \cite{mukherjee1994network}. All anomaly detection approaches below do not distinguish between malicious events and faulty events --- and none of them is able to detect our attacks.

Sommer et al. \cite{sommer} discuss the surprising imbalance between the extensive amount of research on machine learning-based anomaly detection versus the lack of operational deployments of such systems. One of the reasons for that, by the authors, is that the machine learning anomaly detection systems are lacking the ability to bypass the ``semantic gap'': The system ``understands'' that an abnormal activity has occurred, but it cannot produce a message that will elaborate, helping the operator differentiate between an abnormal activity and an attack.

\subsubsection{Network--Aware Detection.}
Basic anomaly detection models for SCADA systems only consider network and OS-level events. Yang et al. \cite{yang2006anomaly} used an Auto Associative Kernel Regression (AAKR) model coupled with the Statistical Probability Ratio Test (SPRT) and applied them on a SCADA system. 
The model used numerous indicators representing network traffic and hardware-operating statistics to predict the `normal' behavior.

Barbosa et al. \cite{barbosa,BarbosaPeriod} analyzed  SCADA traces they collected at two different water treatment and distribution facilities.
They concluded that SCADA traffic presents remarkably regular time series, due to the fact that the majority of the traffic sources generate data in a periodical fashion. They selected only the high energy frequencies for the anomaly detection phase.

\subsubsection{Protocol--Aware Detection.}
More advanced anomaly detection systems rely on deep-packet-inspection and consider the ICS control protocol's meta-data, modeling command sequences, and argument addresses.

Model-based anomaly detection for SCADA systems, and specifically for Modbus traffic, was introduced by Cheung et al. \cite{cheung2007using}. They designed a multi-algorithm intrusion detection appliance for Modbus/TCP with pattern anomaly recognition, Bayesian analysis of TCP headers and stateful protocol monitoring, complemented with customized Snort rules \cite{roesch}. In subsequent work, Valdes et al. \cite{valdes2009communication} incorporated adaptive statistical learning methods into the system to detect for communication patterns among hosts and traffic patterns.

Goldenberg \& Wool \cite{Goldenberg201363} developed a model-based approach (the GW model) for Network anomaly detection based on the normal traffic pattern in Modbus SCADA networks using a Deterministic Finite Automata (DFA) to represent the cyclic traffic. The SCADA messages are modeled both in isolation and also by their sequence order. Subsequently, Kleinmann et al. \cite{KleinmannWjdfsl,kleinmann2015statechart} demonstrated that a similar methodology is successful also in SCADA systems running the Siemens S7 protocol.

Caselli et al. \cite{Caselli} proposed a methodology to model sequences of SCADA protocol messages as Discrete Time Markov Chains (DTMCs).
Based on data from three different Dutch utilities the authors found that only 35\%-75\% of the possible transitions in the DTMC were observed.  This strengthens the observations of \cite{barbosa,Goldenberg201363,KleinmannWjdfsl} of a substantial
sequentiality in the SCADA communications. However, unlike \cite{Goldenberg201363,KleinmannWjdfsl} they did not observe clear cyclic message patterns. The authors hypothesized that the difficulties in finding clear sequences is due to the presence of several threads in the HMI's operating system that multiplex requests on the same TCP stream.

Kleinmann et al. \cite{kleinmann2015statechart,Kleinmann:2016:ACS:2994487.2994490,KleinmannWool2017}
 introduced a modeling approach for multiplexed SCADA streams, using {\em Statechart DFAs}: the {\em Statechart} includes multiple DFAs, one per cyclic pattern. Each DFA is built using the learning stage of the GW model. Following this model, incoming traffic is de-multiplexed into sub-channels and sent to the respective DFAs.

\subsubsection{Process--Aware Detection.}
These anomaly detection methods are based on process invariants: mathematical relationships among physical properties of the process controlled by the PLCs. Several publications \cite{cardenas2011attacks,liu2011false,urbina2016limiting} explain that an IDS that models only the protocol meta-data (commands and arguments) is not sufficient, since attacks can be mounted using legitimate control commands, but with attacker-selected data values. To combat such attacks they suggest modeling both the physical process and the continuous control function. Based on measurements of the state of the process, the models predict the control's response and its effect on the state, and flag deviations from the predicted state.

Fovino et al. \cite{fovino2010modbus} use detailed knowledge of the industrial process' control to generate a system virtual image representing the PLCs of a monitored system. The virtual image is updated using a periodic active synchronization procedure and via a feed generated by the intrusion detection system (i.e., known intrusion signatures).

Sa et al. \cite{sa2016covert} proposed a covert attack against ICS for service degradation, which is planned based on observation how the physical system behaves. Their simulation results demonstrated that attack is able to affect, in a covert and accurate way, the physical behavior of an ICS. They argue that an approach regarding to the covertness of attacks on ICS must be analyzed from two aspects simultaneously: the physical and the cybernetic aspects. Properties of the physical process can be used to predict and then confirm that the control commands sent to the field were executed correctly and that the information coming from sensors is consistent with the expected behavior of the system.

Hadziosmanovic et al. \cite{hadziosmanovic2011} used the logs generated by the control application running on the HMI to detect anomalous patterns of user actions on process control application. The focus of this work was on the threats that can be triggered by a single user action. The authors acknowledged that ``an attacker could manipulate logs by sending false data to the control application''. This model is also susceptible to replay attacks.

Lin et al. \cite{lin2013semantic} combine system knowledge of both the control network (extracting control commands from SCADA network packets) and the physical infrastructure in power grid (obtaining measurements from sensors in substations) to help IDS to estimate execution consequences of control commands, thus to reveal attacker's malicious intentions. The authors claimed that their semantic analysis provides reliable detection of malicious commands with a small amount of analysis time.

Erez et al.\ \cite{ew15-conf} developed an anomaly detection system that detects irregular changes in SCADA control registers' values. The system is based on an automatic classifier that identifies several classes of PLC registers (Sensor, Counter and Constant registers). Parameterized behavior models were built for each class. In its learning phase, the system instantiates the model for each register. During the enforcement phase the system detects deviations from the model.

Mo et al. \cite{mo2012cyber} as well as Pasqualetti et al. \cite{pasqualetti2013attack} investigated the detection and prevention of deception and replay attacks. They concluded that certain types of attacks are undetectable by using their attack models.

\section{Conclusions and Counter Measures}
This work presented a class of semantic network-based attacks against SCADA systems which are undetectable by both protocol--aware and process--aware anomaly detection. After hijacking the communication channels between the HMI and PLCs, our attacks cause the HMI to present a fake view of the industrial process, deceiving the human operator into taking manual actions. Our most advanced attack also manipulates the operator's commands reversing their semantic meaning while causing the HMI to present a view that is consistent with the attempted operator directions. The attacks are totally stealthy since the message sizes and timing, the command sequences, and the data values of the ICS's state all remain legitimate. They appear as natural fault conditions that the SCADA system is designed for, and the human operator is trained to handle.

We implemented and tested several attack scenarios in the realistic test lab of our local electric company. We developed a real-time security assessment tool, that can simultaneously manipulate the communication to multiple PLCs, and cause the HMI to display a coherent system--wide fake view. Our tool is configured with a new IAML language we designed. Our multi-stage semantic attacks all successfully fooled the operator and brought the system to states of blackout and possible equipment damage.

We argue that current intrusion detection and anomaly detection systems are fundamentally unable to detect the stealthy deception attacks we described. In fact, once the attacker is positioned as MITM, the traffic at both the HMI and the PLC sides looks perfectly normal---because it is perfectly normal. In our opinion the only real countermeasure against such attacks is to secure the communication channel via cryptographic means. E.g., by adding data integrity protections such as digital signatures or message authentication codes to block the attacker's ability to modify packets. Nevertheless, we believe that ongoing research into anomaly detection for ICS is still very valuable: with such systems in place, the attacker is restricted to {\em only} mount super-stealthy deception  attacks like ours, and cannot mount simpler and more direct attacks without risk of detection.

\bibliographystyle{splncs03}
\bibliography{amitbib}
\vspace{-0.9em}
\begin{figure*}[!b]
\vspace{-1em}
{\bf Appendix: Example of an IAML script}
\centering

 \lstset{
    language=xml,
    tabsize=2,
    firstnumber=1,
    caption=The IAML script of the multistage attack on a PLC,
    label=lst:IamlAttackMultistage1,
    frame=shadowbox,
    xleftmargin=15pt,
    framexleftmargin=15pt,
    keywordstyle=\color{blue}\bf,
    commentstyle=\color{OliveGreen},
    stringstyle=\color{red},
    numbers=left,
    numberstyle=\tiny,
    numbersep=5pt,
    breaklines=true,
    showstringspaces=false,
    lineskip={-1.5pt},
    basicstyle=\footnotesize,
    basewidth = {.55em,0.55em},
    emph={food,name,price},emphstyle={\color{magenta}}}
\begin{lstlisting}
<?xml version="1.0" encoding="utf-8"?>
<IAML protocol="MODBUS">
  <Change PacketToChange="REQUEST">
    <Query>
      <QueryEntry Key="TYPE" Value="REQUEST"/>
      <QueryEntry Key="FUNCTION" Value="WRITE_MULTIPLE_REGISTERS"/>
      <QueryEntry Key="ADDRESS" Value="129"/>
      <QueryEntry Key="WORD_COUNT" Value="1"/>
    </Query>
    <NewValues>
      <NewValueEntry Key="DATA" Value="41-x"/>
    </NewValues>
  </Change>
  <Change PacketToChange="RESPONSE">
    <Query>
      <QueryEntry Key="TYPE" Value="REQUEST"/>
      <QueryEntry Key="FUNCTION" Value="READ_HOLDING_REGISTERS"/>
      <QueryEntry Key="ADDRESS" Value="129"/>
      <QueryEntry Key="WORD_COUNT" Value="1"/>
    </Query>
    <NewValues>
      <NewValueEntry Key="GLOBAL_STAGE" Value="1"/>
      <NewValueEntry Key="DATA" Value="41-x"/>
    </NewValues>
  </Change>
  <Change PacketToChange="RESPONSE">
    <Query>
      <QueryEntry Key="GLOBAL_STAGE" Value="0"/>
      <QueryEntry Key="TYPE" Value="REQUEST"/>
      <QueryEntry Key="FUNCTION" Value="READ_HOLDING_REGISTERS"/>
      <QueryEntry Key="ADDRESS" Value="129"/>
      <QueryEntry Key="WORD_COUNT" Value="4"/>
    </Query>
    <NewValues>
      <NewValueEntry Key="DATA" Value="0,0,0,x"/>
    </NewValues>
  </Change>
  <Change PacketToChange="RESPONSE">
    <Query>
      <QueryEntry Key="GLOBAL_STAGE" Value="0"/>
      <QueryEntry Key="TYPE" Value="REQUEST"/>
      <QueryEntry Key="FUNCTION" Value="READ_HOLDING_REGISTERS"/>
      <QueryEntry Key="ADDRESS" Value="129"/>
      <QueryEntry Key="WORD_COUNT" Value="16"/>
    </Query>
    <NewValues>
      <NewValueEntry Key="DATA" Value="0,0,0,x,x,x,x,x,x,x,x,x,x,x,x,x"/>
    </NewValues>
  </Change>
\end{lstlisting}
\end{figure*}


\begin{figure*}[t]
\vspace{-1em}
\centering
 \lstset{
    language=xml,
    tabsize=2,
    firstnumber=50,
    frame=shadowbox,
    rulesepcolor=\color{gray},
    xleftmargin=15pt,
    framexleftmargin=15pt,
    keywordstyle=\color{blue}\bf,
    commentstyle=\color{OliveGreen},
    stringstyle=\color{red},
    numbers=left,
    numberstyle=\tiny,
    numbersep=5pt,
    breaklines=true,
    showstringspaces=false,
    lineskip={-1.5pt},
    basicstyle=\footnotesize,
    basewidth = {.55em,0.55em},
    emph={food,name,price},emphstyle={\color{magenta}}}
\begin{lstlisting}
  <Change PacketToChange="RESPONSE">
    <Query>
      <QueryEntry Key="PLC_IP" Value="172.27.21.35"/>
      <QueryEntry Key="GLOBAL_STAGE" Value="1"/>
      <QueryEntry Key="TYPE" Value="REQUEST"/>
      <QueryEntry Key="FUNCTION" Value="READ_HOLDING_REGISTERS"/>
      <QueryEntry Key="ADDRESS" Value="129"/>
      <QueryEntry Key="WORD_COUNT" Value="4"/>
    </Query>
    <NewValues>
      <NewValueEntry Key="DATA" Value="0,12274-x,2334-x,432-x"/>
    </NewValues>
  </Change>
  <Change PacketToChange="RESPONSE">
    <Query>
      <QueryEntry Key="PLC_IP" Value="172.27.21.35"/>
      <QueryEntry Key="GLOBAL_STAGE" Value="1"/>
      <QueryEntry Key="TYPE" Value="REQUEST"/>
      <QueryEntry Key="FUNCTION" Value="READ_HOLDING_REGISTERS"/>
      <QueryEntry Key="ADDRESS" Value="129"/>
      <QueryEntry Key="WORD_COUNT" Value="16"/>
    </Query>
    <NewValues>
      <NewValueEntry Key="DATA" Value="0,12274-x,2334-x,432-x,x,x,x,x,x,x,x,x,x,x,x,x"/>
    </NewValues>
  </Change>
  </IAML>
\end{lstlisting}
\end{figure*}

\end{document}